\documentclass[aps,prl,twocolumn,showpacs]{revtex4}
\usepackage{times}
\usepackage{bm}
\usepackage{graphicx}

\newcommand{\smeq}{\! = \!}

\newcommand{\smmi}{\! - \!}

\newcommand{\ve}{\varepsilon}
\newcommand{\kf}{k_{\text{F}}}
\newcommand{\Ef}{E_{\text{F}}}
\newcommand{\kt}{k_{\text{B}}T}
\newcommand{\be}{\begin{equation}}
\newcommand{\ee}{\end{equation}}
\newcommand{\bea}{\begin{eqnarray}}
\newcommand{\eea}{\end{eqnarray}}

\begin{document}

\title{Tuning the Non-local Spin-Spin Interaction between Quantum Dots with a Magnetic
Field}
\author{Gonzalo Usaj}
\affiliation{Instituto Balseiro and Centro At\'{o}mico Bariloche, Comisi\'{o}n 
Nacional de Energ\'{\i}a At\'{o}mica, (8400) San Carlos de Bariloche, Argentina.}
\author{P. Lustemberg}
\affiliation{Instituto Balseiro and Centro At\'{o}mico Bariloche, Comisi\'{o}n 
Nacional de Energ\'{\i}a At\'{o}mica, (8400) San Carlos de Bariloche, Argentina.}
\author{C. A. Balseiro}
\affiliation{Instituto Balseiro and Centro 
At\'{o}mico Bariloche, Comisi\'{o}n Nacional de Energ\'{\i}a At\'{o}mica, (8400) San 
Carlos de Bariloche, Argentina.}
\date{2 July 2004}

\begin{abstract}
We describe a device where the non-local spin-spin interaction between
quantum dots (QDs) can be turned on and off with a small magnetic
field. The setup consists of two QDs at the edge of two
two-dimensional electron gases (2DEGs). The QDs' spins are coupled
through a RKKY-like interaction mediated by the electrons in the
2DEGs. A magnetic field $B_z$ perpendicular to the plane of the 2DEG
is used as a tuning parameter. When the cyclotron radius is
commensurate with the interdot distance, the spin-spin interaction is
amplified by a few orders of magnitude. The sign of the interaction is
controlled by finely tuning $B_z$. Our setup allows for several dots
to be coupled in a linear arrangement and it is not restricted to
nearest-neighbors interaction. 
\end{abstract}
\maketitle

Quantum information processing requires control and operation of interacting
quantum mechanical objects \cite{QCbook}. One possibility is to produce
systems with localized spins in atomic impurities, molecules or quantum dots
(QDs) and manipulate the spin-spin interaction by engineering the electronic
wave functions of the surrounding material \cite{LossD98,BurkardLD99,CorreaHB02}. This would
allow for the non-local control of spins opening new possibilities
for the fast developing field of spintronics \cite{Spintronicsbook}.
An important step in this direction was reported very recently by
Craig \textit{et al.} \cite{CraigTLMHG04}, who coupled two QDs through
a confined 2DEG (a larger QD) and controlled the magnitude of the
interaction by closing or opening up the QDs. Besides its relevance
for spintronics, this experiment also opens up the possibility to
study the interplay between two competing many-body effects: the Kondo
effect and the RKKY-interaction
\cite{GalkinBA04,VavilovG04,SimonLO04}.

In this work, we propose a different device where the magnitude (and sign) of the spin-spin interaction between two QDs can
be tuned by a external field. Our setup consists of two QDs at the edge of two semi-infinite
two dimensional electron gases (2DEGs). When each dot is gated to have
an odd number of electrons, and therefore a total spin $ \frac{1}{2}
$, they interact through the polarization of the 2DEG as in the usual case described by the RKKY interaction \cite{RKKY}.
We show that this interaction can be controlled by applying a small
magnetic field perpendicular to the plane of the 2DEG. The effect
relies on the existence of edge states. 
These are responsible for the transverse focusing of electrons injected from a
point contact \cite{BeenakkerH91,PotokFMU02,UsajB04_focusing} or QD \cite{PotokFMUHG03}. The control mechanism is based on the possibility of
focusing the electrons that interact with one QD onto the other by
the action of the external field. When the cyclotron radius is commensurate
with the interdot distance, the spin-spin interaction is largely amplified
and may increase a few orders of magnitude. The mechanism can be extended to
many spins and to new geometries that allow for the independent control of
different pair interactions.
 
\begin{figure}[t]
   \centering
   \includegraphics[height=3.5cm,clip]{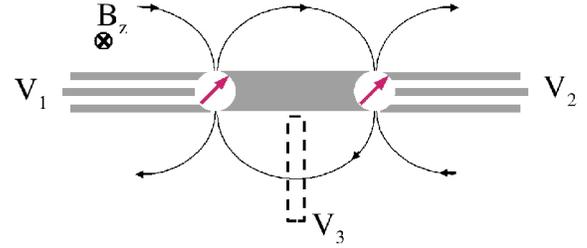}
   \caption{Schematic representation of the proposed device. Both QDs
   are setup to have an odd number of electrons---this is controlled
   by the gate voltages $ V_1 $ and $ V_2 $. The QDs spins are then
   coupled through a RKKY-like interaction mediated by the electrons
   in the two 2DEGs. This interaction can be tuned by a perpendicular
   magnetic field $ B_z $, being maximum when the cyclotron orbit
   matches the distance between the two QD. A third (optional) gate $
   V_3 $ can be used to interrupt one of the electrons' path and
   cancel out the interaction.} 
   \label{scheme}
\end{figure}

Consider the geometry shown in Fig.\ref{scheme}. It
consists of a 2DEG with a small
magnetic field $B_{z}$ perpendicular to the plane containing the carriers.
The 2DEG is divided in two half planes by a set of gate contacts which
are also used to define the QDs as schematically shown in the
figure. The Hamiltonian of the system reads 
 
\begin{equation}
{\hat{H}}\!=\!{\hat{H}}_{\text{QD}}\!+\!{\hat{H}}_{\text{2DEG}}\!+\!{\hat{H}}%
_{\text{T}}\,,  
\label{h}
\end{equation}
where the first term is the Hamiltonian of the QDs
\begin{equation}
{\hat{H}}_{\text{QD}}\!=\!\sum_{\alpha ,\sigma }E_{\alpha \sigma }^{}d_{\alpha
\sigma }^{\dagger }d_{\alpha \sigma }^{}\!+\!U_{\alpha }^{}d_{\alpha \uparrow
}^{\dagger }d_{\alpha \uparrow }^{}d_{\alpha \downarrow }^{\dagger }d_{\alpha
\downarrow }^{}\,.
\end{equation}
Here $\alpha \!=\!1$ and $2$ refer to the left and right QD
respectively, $d_{\alpha \sigma }^{\dagger }$ creates an electron with
spin $\sigma $ and 
energy $E_{\alpha \sigma }$ in the QD labeled by $\alpha $ and $U_{\alpha }$ is the
Coulomb energy defined by the capacitances of the system \cite{AleinerBG02}. 
The single particle energies $E_{\alpha \sigma }$ (measured from the Fermi
energy, $\Ef$) can be varied with a gate voltage $V_{\alpha }$ as shown in Fig.
\ref{scheme}. The second term in Hamiltonian (\ref{h}) describes the electrons in the two
half planes. To describe these 2DEGs, we discretize the space and use a
tight binding model on a square lattice,
\begin{equation}
{\hat{H}}_{\text{2DEG}}\!=\!\sum_{\gamma \mathbf{n}\sigma }\varepsilon
_{\sigma }^{}c_{\gamma \mathbf{n}\sigma }^{\dagger }c_{\gamma \mathbf{n}\sigma
}^{}\!-\!\sum_{<\mathbf{n},\mathbf{m}>\sigma }t_{\mathbf{n}\mathbf{m}}^{}\,c_{\gamma 
\mathbf{n}\sigma }^{\dagger }c_{\gamma \mathbf{m}\sigma }^{}+\mathrm{h.c.}
\end{equation}
where $c_{\gamma \mathbf{n}\sigma }^{\dagger }$ creates an electron with
spin $\sigma $ at site $\mathbf{n}\!=\!(n_{x},n_{y})$ of the upper ($\gamma
\!=\!1$) and lower ($\gamma \!=\!2$) half planes. The hopping matrix element 
$t_{\mathbf{n}\mathbf{m}}$ connects nearest neighbors and includes the
effect of the diamagnetic coupling through the Peierls substitution. To
avoid any zone boundary effects we take a lattice parameter $a_{0}\smeq5$nm much
smaller than the characteristic Fermi wavelength ($\lambda_\mathrm{F}\!\sim\!50$nm)
. We use the Landau gauge
for which $t_{\mathbf{n}(\mathbf{n}\!+\!\widehat{\mathbf{x}}\mathbf{)}%
}\!=\!te^{-in_{y}2\pi\phi /\phi _{0}}$ and $t_{\mathbf{n}(\mathbf{n}\!+\!%
\widehat{\mathbf{y}}\mathbf{)}}\!=\!t$, where $\phi \!=\!a_{0}^{2}B_{z}$ is the
magnetic flux per plaquete, $\phi _{0}\!=\!hc/e$ is the flux quantum and
$t\!=\!\hbar ^{2}/2m^{\ast }a_{0}^{2}$ with $m^{\ast }$ is the carriers
effective mass. The third term in the Hamiltonian describes the coupling
between the QD and the 2DEGs
\begin{equation}
{\hat{H}}_{\text{T}}\!=\!-\!\sum_{\alpha \gamma \sigma }t_{\alpha \gamma
}(c_{\gamma \alpha \sigma }^{\dagger }d_{\alpha \sigma }\!+\!d_{\alpha
\sigma }^{\dagger }c_{\gamma \alpha \sigma })\,,
\end{equation}
where $c_{\gamma \alpha \sigma }^{\dagger }\smeq
N_{0}^{-\frac{1}{2}}\sum_n c_{\gamma n \sigma }^{\dagger} $ creates an
electron at the half plane $\gamma $ in a linear combination of $N_0$
sites next to the QD $\alpha $. We are interested in  
the range of magnetic fields that produce a cyclotron radius $r_{c}$ of the
order of the interdot distance $R$, defined as the average distance between the sites connected to dot $1$ and dot $2$.
For these small fields, the Zeeman splitting can be neglected restoring the
spin rotational symmetry. 
In what follows we take $E_{\alpha \uparrow }\!=\!E_{\alpha \downarrow }$. The generalization to the case of a
large Zeeman energy is straightforward.
We assume that the QDs are gated to be in the strong Coulomb blockade
regime, $U\!+\!E_{\alpha }\!\simeq \!-E_{\alpha }$, so that the charge
fluctuations can be eliminated by the Schrieffer-Wolf transformation \cite
{SchriefferW66}. The spin dynamics is then described by a Kondo Hamiltonian
where ${\hat{H}}_{\text{QD}}\!+\!{\hat{H}}_{\text{T}}$ is replaced by \cite
{Hewson}
\begin{equation}
{\hat{H}}_{K}\!=\!\sum_{\alpha }J_{\alpha }\vec{S}_{\alpha }\cdot
(c_{1\alpha \sigma }^{\dagger }\!+\!c_{2\alpha \sigma }^{\dagger })\frac{%
\vec{\sigma}_{\sigma \sigma ^{\prime }}}{2}(c_{1\alpha \sigma ^{\prime
}}^{}\!+\!c_{2\alpha \sigma ^{\prime }}^{})\,,  
\label{hamt}
\end{equation}
where $\vec{S}_{\alpha }$ is the spin operator associated to the QD $\alpha $
and
\begin{equation}
J_{\alpha }\!=\!2|t_{\alpha }|^{2}\left( \frac{1}{U_{\alpha }\!+\!E_{\alpha }%
}\!-\!\frac{1}{E_{\alpha }}\right) \simeq \frac{8|t_{\alpha }|^{2}}{%
U_{\alpha }}\smeq\frac{\Gamma _{\alpha}}{U_{\alpha }}\frac{4}{\pi\rho}\,,
\label{jk}
\end{equation}
with $\Gamma _{\alpha }$ the level width and $\rho $ the local density of states per spin at $\Ef$. $%
\Gamma _{\alpha }$ and $U_{\alpha }$ can be measured by transport experiments \cite{Goldhaber-GordonGKSMM98,vanderWielFFETK00}.
For simplicity, we take $t_{\alpha \gamma }\equiv t_{\alpha }$ and neglect a potential scattering term which is
not relevant for the present work \cite{Hewson}. 
Finally, usual
perturbative procedures give an effective exchange interaction between the
QDs' spins mediated by electrons in the 2DEG. 
The interdot exchange Hamiltonian ${\hat{H}}_{J}$ contains the non-local susceptibility which can be written
in terms of one-particle propagators.
With a negligible Zeeman splitting, the propagators
are spin independent and the Hamiltonian reduces to ${\hat{H}}_{J%
}\!=\!J\vec{S}_{1}\cdot \vec{S}_{2}$ with 
\begin{equation}
J\!=\!-\frac{J_{1}J_{2}}{2\pi }\int d\omega f(\omega )\mathrm{Im}%
[G_{\downarrow }(1,2)G_{\uparrow }(2,1)]\,,
  \label{jrkky}
\end{equation}
where $\mathrm{Im}$ denotes the imaginary part and $G_{\sigma }(\alpha
,\alpha ^{\prime })\!=\!\langle \langle (c_{1\alpha \sigma }\!+\!c_{2\alpha
\sigma }),(c_{1\alpha ^{\prime }\sigma }^{\dagger }\!+\!c_{2\alpha ^{\prime
}\sigma }^{\dagger })\rangle \rangle _{\omega \!+\!\mathrm{i}\eta }$ is the
Fourier transform of the retarded Green function \cite{Zubarev60} and $f(\omega )$
is the Fermi function. In the following we take $\kt\!\ll\!J$---in this regime $f(\omega )\!\simeq\!\Theta(\Ef\smmi\omega)$.
To lowest order in $J_{\alpha }$, the one-particle propagators are
calculated with ${\hat{H}}_{\text{2DEG}}$ only. Then, we have
\begin{equation}
G_{\sigma }(\alpha ,\alpha ^{\prime })\!=\!g_{1\sigma }(\alpha ,\alpha
^{\prime })\!+\!g_{2\sigma }(\alpha ,\alpha ^{\prime })\,,
\label{propaga}
\end{equation}
with $g_{\gamma \sigma }(\alpha ,\alpha ^{\prime })\!=\!\langle \langle
c_{\gamma \alpha \sigma },c_{\gamma \alpha ^{\prime }\sigma }^{\dagger
}\rangle \rangle _{\omega +\mathrm{i}\eta }$. Equation (\ref{jrkky}) makes
evident that, in terms of Feynman diagrams, the effective interaction---or
the non-local susceptibility---is a bubble diagram with a propagator from
dot $1$ to dot $2$ times a propagator from dot $2$ to dot $1$.
In the following, we calculate these propagators numerically (see Ref. [\onlinecite{UsajB04_focusing}] for details). 
\begin{figure}[t]
   \centering
   \includegraphics[height=6.3cm,clip]{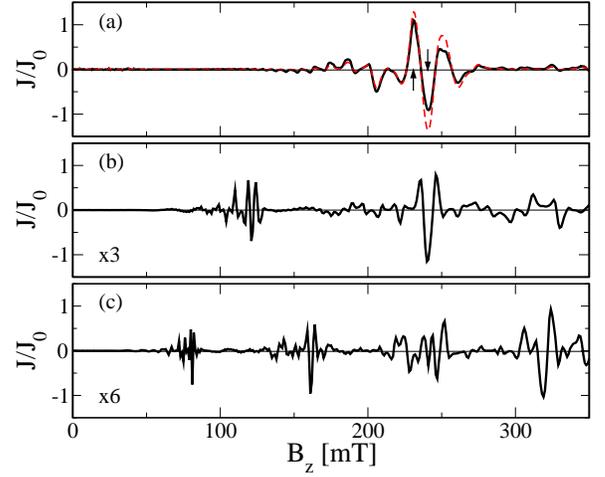}
   \caption{(Color online). Exchange interaction $ J $ as a function
     of the perpendicular magnetic field $ B_z $ for different
     interdot distances $ R\smeq 0.5\, \mu m $ (a), $1\, \mu m $ (b),
     and $ 1.5\, \mu m $ (c) and $N_0\smeq1$ ($N_0\smeq7$ is shown in
     (a) [dashed line]). In the last two cases, $ J $ was multiplied
     by $ 3 $ and $ 6 $ respectively.  
The interaction presents large oscillations when the cyclotron radius
$ r_c $ is such that $ 2r_c\!\simeq\! R/n $ with $ n $ an integer
number. A fine tuning of $ B_z $ allows then to select the
\textit{magnitude} and the \textit{sign} of the exchange interaction. 
}
\label{field}
\end{figure}

The field $B_{z}$ creates edge states that propagate in opposite directions
on opposite sides of the QDs (see Fig. \ref{scheme}). This generates a
right-left asymmetry on \textit{each} half plane, \textit{ie.} $g_{\gamma \sigma }\left(
1,2\right) \neq $ $g_{\gamma \sigma }\left( 2,1\right) $. In general, with $%
B_{z}\neq 0$, one of these propagators is very small. In
fact, with a single half plane, the backscattering of the edge states is
strongly suppressed and they do not contribute to the non-local
susceptibility. In the proposed geometry, however, tunneling between the two
half planes described by Hamiltonian (\ref{hamt}) creates a channel for backscattering 
and the product $g_{1\sigma }\left( 1,2\right) $ $g_{2\sigma
}\left( 2,1\right) $ does contribute to the effective interdot coupling. For
this reason, an effective non-vanishing and controllable spin-spin
interaction requires a configuration with the two half planes.

\begin{figure}[t]
   \centering
   \includegraphics[height=6.4cm,clip]{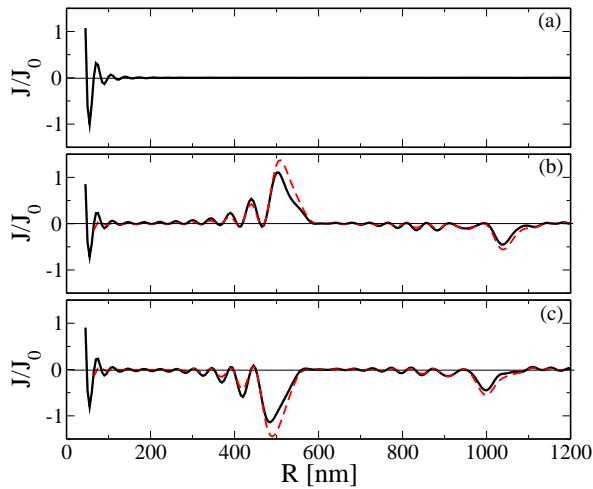}
   \caption{(Color online). Exchange interaction as a function of the interdot distance $ R $ for $ B_z\smeq 0 $ (a), $ 231$mT (b) and $ 241$mT (c). These values of $ B_z $ are indicated in Fig. \ref{field}a with arrows. The magnitude of $ J $ increases a few orders of magnitude when $ R\!\simeq\! 2r_c\!\simeq\!500\,nm$ while its sign depends on the precise value of $ B_z $. The solid and dashed lines correspond to $N_0\smeq1$ and $7$, respectively.}
   \label{distance}
\end{figure}

Results for the exchange coupling $J$ are shown in Figure \ref{field} and 
\ref{distance} normalized by $J_0=J(R\!\sim\!\lambda_F,B_z\smeq0$).
We used parameters typical of GaAs systems---an electron
density of $n\!=\!1.5\times 10^{11}\,cm^{-2}$, that corresponds to a Fermi
energy $E_{F}\!=\!5meV$\ measured from the bottom of the conduction band and
a Fermi momentum $\kf \simeq 0.1\,nm^{-1}$.
Figure 2 shows  $J$ as a function of $B_{z}$ for interdot distances $R\!=\!0.5$, $1
$ and $1.5\mu m$. In all cases, the interaction presents large oscillations
whenever the magnetic field is such that twice the cyclotron radius $r_{c}\!=\!\hbar
ck_{F}/eB_{z}$ is commensurable with $R$. We refer to this fields as the
focusing fields since in this situation electrons that interact with one
dot are focused into the other by the action of the field. As $R$ increases the characteristic period of the oscillations and
their amplitude decreases (the nature of these oscillations is discussed below). 
The exchange coupling $J$ as a function of the interdot distance is shown
in Fig. \ref{distance} for fixed fields. For $B_{z}\!=\!0$ the dominant contribution is $J\propto \!\cos (2k_{\text{F}%
}R)/(\kf R)^{4}$, in contrast to the conventional $2$D behavior $J\propto \!\cos (2\kf R)/(\kf R)^{2}$.
 The power law decay $R^{-4}$ is due to
the structure of the states near the edge of the 2DEG. 
Namely, $\rho$ depends linearly on energy (it is constant in bulk). Since in a semiclassical picture the contribution to the propagator
Eq. (\ref{propaga}) arises only from the classical trajectories near the boundary, one could argue that the effective density
is $\rho(\ve)\!\propto\!\ve$, so this case ``mimics'' 
a $4$D one (thus the $R^{-4}$ decay). This is confirmed by both a quantum and a semiclassical analytical calculation.     
For $B_{z}\!\neq \!0$, a large amplification of $J$ is observed for
distances such that $R\!=\!n2r_{c}$, where $n\!$ is an integer. Comparison
of Figs. \ref{distance}a and \ref{distance}b shows that at distances of
the order of $0.5\mu m$, an increase of the field from zero to its focusing
value increases the coupling by more than two
orders of magnitude (notice $J/J_0\!\sim\!1$ for $n\smeq1$). Additionally, the sign of the interaction is controlled by finely tuning the magnetic field around the focusing value (see arrows in Fig. \ref{field}a). Note that a larger $N_0$ (tunneling region) enhances the effect. This is due to a reduction of diffraction effects, which leads to a better definition of the classical cyclotron orbit \cite{UsajB04_focusing}. 
\begin{figure}[t]
   \centering
   \includegraphics[height=7.4cm,clip]{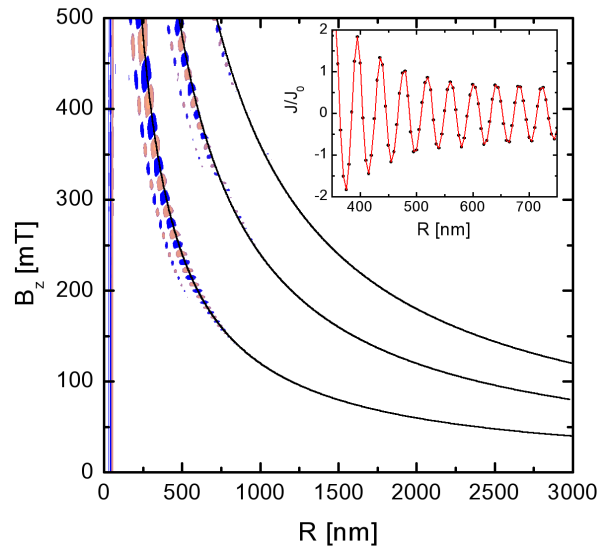}
   \caption{ (Color online). Density plot of the exchange interaction $ J $ as a function of the perpendicular magnetic 
field $ B_z $ and the interdot distance $ R $. The solid lines correspond to different focusing conditions with $ n\smeq1,2,3 $ (cf. Eq. (\ref{ff})). Note that the magnitude of the interaction is large along those lines while its sign changes (dark and light areas). Inset: $ J $ as a function of $ R $ along the first hyperbola. The solid line is a fitting to Eq. (\ref{jcos}).}
   \label{density}
\end{figure}

All these results can be put together in a single density plot as shown in
Fig. \ref{density}. The lines are hyperbolas corresponding to different focusing fields 
\be
B_{z}^{(n)}\!=\!2n(\hbar c\kf /eR)\!=\!2n\phi _{0}/\lambda _{\text{F}}R\,.
\label{ff}
\ee
Along these lines, \textit{ie.} when field and distance are
simultaneously varied to keep the focusing condition, the exchange
coupling oscillates with large amplitude. A simple fitting of the
numerical results shows that the dominant contribution to $ J $
along the hyperbolas is given by 
\begin{equation}
J\!\propto\! \frac{\cos (\kf \pi R/2\!+\!\varphi )}{R^{2}}\,.
\label{jcos}
\end{equation}
Note that the interaction decays as $R^{-2}$ as in the usual $2$D case---this is consistent with the semiclassical picture since now the classical trajectories are not restricted to be close to the boundary.
The argument
of the cosine can be written as $\kf R_{\text{eff}}$ where $R_{\text{eff}}\smeq\pi R/2$ 
is the length of the classical trajectory (see Fig.\ref{scheme}). The period of the oscillations is then equal to the period at which
the Landau levels cross $\Ef$. In fact, we have 
$\kf R_{\text{eff}}\smeq\!2\pi R_{\text{eff}}/\lambda _{\text{F}}\smeq2\pi \Ef/\hbar \omega
_{c}$. These oscillations of the non-local susceptibility are the analog of
the de Haas-van Alphen oscillations of the magnetization. The complex oscillating pattern observed in Fig. \ref{field} corresponds to a cut in Fig.\ref{density} along a vertical line. The inset in Fig.\ref{density}
shows the coupling $J$ when the field and the
interdot distance are varied along the first hyperbola ($n\smeq1$). 
The solid line is a fitting to Eq. (\ref{jcos}). 

\begin{figure}[t]
   \centering
   \includegraphics[height=5.8cm,clip]{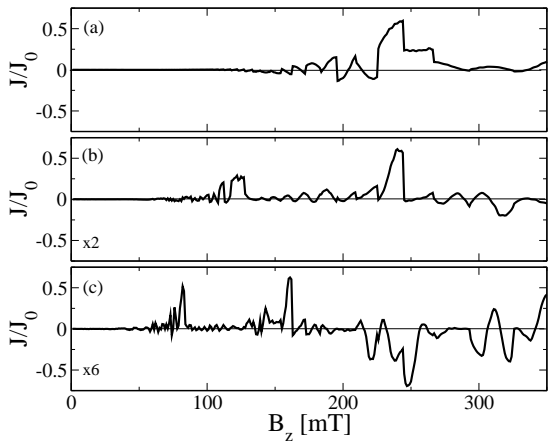}
   \caption{Exchange coupling as a function of $ B_z $ with the same parameters as in Fig. \ref{field} but 
with the chemical potential pinned at the partially filled Landau Level. Note that the sign of the interaction tends to be preserved around each focusing field.}
   \label{neutral}
\end{figure}

So far we have presented results obtained by fixing the chemical potential. 
However, in a 2DEG with a fixed charge density, $E_{\text{F}}$ is pinned at the energy of
the partially filled Landau level and presents periodic jumps when plotted
as a function of $1/B_{z}$. Figure \ref{neutral} shows the coupling $J$
versus the external field for constant electron density at the bulk of the
2DEG. Now again, the spin-spin interaction presents a large enhancement at
the focusing fields. There are, however, some differences with the previous
case: as the external field is varied around the focusing values, the sign
of the interdot interactions tends to be preserved. Both ferromagnetic and
antiferromagnetic couplings are obtained. The dominant sign of  $J$ at the
different focusing field depends on the parameters, in particular on the
particle density. 
The jumps of the Fermi energy overestimate some charge redistribution at the
edges. Including the electron-electron interactions in a self-consistent
approximation would tend to preserve local charge neutrality. This may
generate an intermediate situation where the effective coupling changes sign
as $B_{z}$ sweeps the focusing values. 

It is worth pointing out that in the presence of a magnetic field, the 
exchange interaction between two
spins in the bulk of a
2DEG also shows some structure: there is a small enhancement of $J$ for $R\!\sim \!2r_{c}$ 
while it decays exponentially for $R\!\geq \!2r_{c}$. 
In the proposed geometry, however,  the focusing effect produces
an amplification of $J$ much larger than what is obtained in an homogeneous
2DEG. Moreover, with present technologies, it is possible to built a
device like the one schematically shown in Fig. \ref{scheme}, where the contacts used
to control the QD parameters ($E_{\alpha }$ and $t_{\alpha }$) are also used to
divide the 2DEG in two halves. An interesting advantage of our setup is that it allows to change the magnitude 
of the exchange coupling between QDs without modifying their coupling to the 2DEG. 
Therefore, transport measurements through the coupled and decoupled system can be easily compared.

Since the exchange mechanism requires the two half planes, interrupting the particle
propagation in one of them decouples the QDs. This provides an alternative way
to act on the effective coupling, which can be implemented with a
gate voltage on a side contact indicated as $V_{3}$ in Fig. \ref{scheme}. 
Also, three or more QDs could be built along the central gate with the same or
different interdot distances, allowing for a variety of alternatives in
which, with the help of side contacts and the external magnetic field $ B_z $, the different couplings could be
varied in sign and magnitude.  It is important to emphasize that the
interaction between QDs is not restricted to nearest neighbors 

In summary, we showed that two QDs at the edge of a 2DEG interact with an
exchange coupling $J$ that can be controlled with a
small magnetic field perpendicular to the 2DEG. When the
cyclotron radius $r_{c}$ becomes commensurable with the interdot distance $R$
there is a large amplification of the interdot interaction.
This condition, $2nr_{c}\! = \! R$, defines the focusing fields. As the
external field is varied around this values, the enhanced interaction changes sign
allowing for a fine tuning of a ferromagnetic or an antiferromagnetic coupling.

This work was partially supported by ANPCyT Grant N$^{\mathrm{o}}$ 99 3-6343, CNRS-PICS
Collaboration Program between France and Argentina, and Fundaci\'on
Antorchas, Grant 14169/21. GU acknowledge financial support from CONICET.

\end{document}